\documentclass[prl,12pt,tightenlines]{revtex4}
\usepackage{latexsym}
\usepackage{graphicx}
\usepackage{amsmath}
\usepackage{amsbsy}
\usepackage{amsthm}
\usepackage{bbm}
\usepackage{bm}

\newtheorem{lem}{Lemma}

\begin{document}

\title{Geometric entanglement witnesses and bound entanglement}
\author{Reinhold A. Bertlmann} \email{reinhold.bertlmann@univie.ac.at}
\author{Philipp Krammer} \email{philipp.krammer@univie.ac.at}
\affiliation{Faculty of Physics, University of Vienna, Boltzmanngasse 5, A-1090 Vienna, Austria}

\begin{abstract}

We study entanglement witnesses that can be constructed with regard to the geometrical structure
of the Hilbert-Schmidt space, i.e. we present how to use these witnesses in the context of
quantifying entanglement and the detection of bound entangled states. We give examples for a
particular three-parameter family of states that are part of the \emph{magic simplex} of
two-qutrit states.

\end{abstract}

\maketitle

The determination whether a general quantum state is entangled or not is of utmost
importance in quantum information. States of composite systems can be classified in
Hilbert space as either entangled or separable. The geometric structure of entanglement
is of high interest, particularly in higher dimensions. For $2 \times 2$ bipartite
qubits, the geometry is highly symmetric and very well known. In higher dimensions, like
$3 \times 3$ two-qutrit states, the structure of the corresponding Hilbert space is more
complicated and less known. New phenomena like bound entanglement occur. In principle,
all entangled states can be detected by the entanglement witness procedure. We
construct in this Brief Report entanglement witnesses with regard to the geometric
structure of the Hilbert-Schmidt space and explore a specific class of states, the
three-parameter family of states. We discover by our method new regions of bound
entanglement.

Let us recall some basic definitions we need in our discussion. We consider a
Hilbert-Schmidt space ${\cal A} = {\cal A}_1 \otimes {\cal A}_2 \otimes \ \ldots \
\otimes {\cal A}_n$ of operators acting on the Hilbert space of $n$ composite quantum
systems - it is of dimension $d_1 \times d_2 \times \ \ldots \ \times d_n$ and $D := d_1
d_2 \ldots d_n$. In Ref.~\cite{witte99} a new geometric entanglement measure (which we
call \emph{Hilbert-Schmidt measure}) based on the Hilbert-Schmidt distance between
density matrices is presented -- and discussed in Ref.~\cite{ozawa00} -- that is an
instance of a \emph{distance} measure (see Refs.~\cite{vedral97, vedral98}). The
entanglement of a state $\rho$ can be quantified (or ``measured'') via the minimal
Hilbert-Schmidt distance of the state to the convex and compact set of separable
(disentangled) states $S$:
\begin{equation} \label{defhs}
    D(\rho) := \min_{\sigma \in S} \| \sigma - \rho \| \,.
\end{equation}
It is defined on the Hilbert-Schmidt metric with a scalar product between operators that
are elements of the Hilbert-Schmidt space ${\cal A}$: $\langle A, B \rangle :=
\textnormal{Tr} A^\dag B, \; A,B \in {\cal A}$ and the norm $\| A \| := \sqrt{\langle A,
A \rangle}\,$. Of course these definitions apply to density matrices (states of the
quantum system), since they are operators of ${\cal A}$ with the properties $\rho^\dag =
\rho$, Tr$\rho = 1$ and $\rho \geq 0$ (positive semidefinite operators).

Because the norm is continuous and the set of separable states $S$ is compact the minimum
in Eq.~\eqref{defhs} is attained for some separable state $\sigma_0$, $\min_{\sigma \in
S} \| \sigma - \rho \| = \| \sigma_0 - \rho \|\,$, which we call the \emph{nearest
separable state} to $\rho$. Clearly if $\rho \in S$ then $\sigma_0 = \rho$ and $D(\rho) =
0$. If $ \rho$ is entangled, $\rho_{\rm{ent}}$, then $D(\rho) > 0$ and $\sigma_0$ lies on
the boundary of the set $S$. It is shown in Refs.~\cite{pittenger02, bertlmann02,
pittenger03, bertlmann05} that an operator
\begin{equation} \label{optoperator}
A_{\rm{opt}} = \sigma_0 - \rho_{\rm{ent}} - \left\langle \sigma_0 ,
    \sigma_0 - \rho_{\rm{ent}} \right\rangle \mathbbm{1}_{\rm{D}}
\end{equation}
defines a hyperplane including the state $\sigma_0$ that is tangent to $S$, and it has the
properties
\begin{align}
    \left\langle \rho_{\rm{ent}},A_{\rm{opt}} \right\rangle = \textnormal{Tr}\, \rho_{\rm{ent}}
    A_{\rm{opt}} & < 0 \,, \label{defentwitent} \\
    \left\langle \sigma,A_{\rm{opt}} \right\rangle = \textnormal{Tr}\, \sigma A_{\rm{opt}}
    & \geq 0 \quad \forall \sigma \in S \,. \label{defentwitsep}
\end{align}
It is clearly Hermitian and thus is an \emph{entanglement witness} for the state
$\rho_{\rm{ent}}$ \cite{horodecki96, terhal00}. Moreover, since the hyperplane is tangent
to $S$ (Tr$\sigma_0 A_{\rm{opt}} = 0$) the operator $A_{\rm{opt}}$ \eqref{optoperator} is
called \emph{optimal entanglement witness}. We call any entanglement witness that is
constructed in the way of Eq.~\eqref{optoperator} -- with any two states, not necessarily
the nearest separable state -- a \emph{geometric entanglement witness}.

The crucial point lies in finding the nearest separable state: Once found, we can both quantify
the entanglement of a state and construct an (even optimal) entanglement witness. But finding the
nearest separable state is a hard task. A ``guess-method'', that is a method to check if a good
guess for the nearest separable state is indeed right, is presented in Ref.~\cite{bertlmann05}.
Another way is to look for the nearest state that is positive under partial transposition (PPT)
first. A method that finds the \emph{nearest PPT state} in many cases is presented in
Ref.~\cite{verstraete02}.

The PPT-criterion \cite{peres96, horodecki96} is a necessary criterion for separability
(sufficient for $2 \times 2$ or $2 \times 3$ dimensional Hilbert spaces): A separable state has to
stay positive semidefinite under partial transposition -- i.e. transposition in only one
subsystem. Thus if a density matrix becomes indefinite under partial transposition, i.e. one or
more eigenvalues are negative, it has to be entangled and we call it a \emph{NPT entangled state}.
But there exist entangled states that remain positive semidefinite -- \emph{PPT entangled states}
-- these are called \emph{bound entangled states}, since they cannot be distilled to a maximally
entangled state \cite{horodecki97a, horodecki98}.

The set of all PPT states $P$ is convex and compact and contains the set of separable
states. Thus the nearest separable state $\sigma_0$ can be replaced by the nearest PPT
state $\tau_0$ for which the minimal distance to the set of PPT states is attained,
$\min_{\tau \in P} \| \tau - \rho \| = \| \tau_0 - \rho \| \,$. If $\rho$ is a NPT
entangled state $\rho_{\rm{NPT}}$ and $\tau_0$ the nearest PPT state, then the operator
\begin{equation} \label{operator}
    A_{\rm{PPT}} := \tau_0 - \rho_{\rm{NPT}} - \langle \tau_0, \tau_0 - \rho_{\rm{NPT}}
    \rangle \mathbbm{1}_{\rm{D}}
\end{equation}
defines a tangent hyperplane to the set $P$ for the same geometric reasons as operator
\eqref{optoperator} and has to be an entanglement witness since $P \supset S$. In principle the
entanglement of $\rho_{\rm{NPT}}$ can be measured in experiments that should verify
Tr$A_{\rm{PPT}} \rho_{\rm{NPT}} < 0$. If the state $\tau_0$ is separable, it has to be the nearest
separable state $\sigma_0$ since the operator \eqref{operator} defines a tangent hyperplane to the
set of separable states. Therefore in this case $A_{\rm{PPT}}$ is an optimal entanglement witness,
$A_{\rm{PPT}} = A_{\rm{opt}}$, and the Hilbert-Schmidt measure of entanglement can be readily
obtained. If $\tau_0$ is not separable, that is PPT and entangled, it has to be a bound entangled
state.

Unfortunately it is not trivial to check if the state $\tau_0$ is separable or not. It is hard to
find evidence of separability, but it might be easier to reveal bound entanglement, not only for
the state $\tau_0$ but for a whole family of states. A method to detect bound entangled states we
are going to present.

\begin{figure}
  \includegraphics[width=0.40\textwidth]{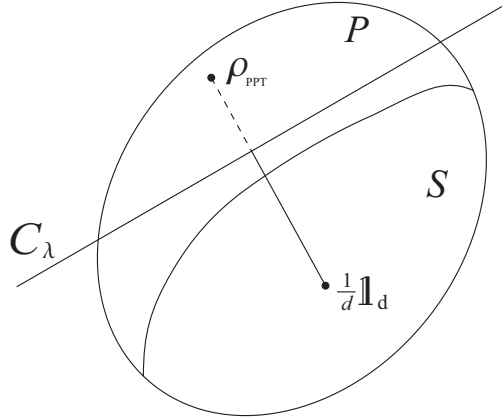}\\
  \caption{Sketch of the presented method to detect bound entanglement with the geometric
    entanglement witness $C_\lambda$. The dashed line indicates the detected bound entangled states
    $\rho_\lambda$ ($\lambda_{\rm{min}} < \lambda \leq 1$), states $\rho_\lambda$ with
    $0 < \lambda \leq \lambda_{\rm{min}}$ can be separable or bound entangled
    (straight line).}\label{figgeneral}
\end{figure}

Consider any PPT state $\rho_{\rm{PPT}}$ and the family of states $\rho_\lambda$ that lie
on the line between $\rho_{\rm{PPT}}$ and the maximally mixed (and of course separable)
state $\frac{1}{D} \mathbbm{1}_{\rm{D}}$,
\begin{equation}
    \rho_\lambda := \lambda \,\rho_{\rm{PPT}} +  \frac{(1-\lambda)}{D} \mathbbm{1}_{\rm{D}} \,.
\end{equation}
We can construct an operator $C_\lambda$ in the following way:
\begin{equation} \label{cbe}
    C_\lambda = \rho_\lambda - \rho_{\rm{PPT}} - \langle \rho_\lambda , \rho_\lambda -
    \rho_{\rm{PPT}} \rangle \mathbbm{1}_{\rm{D}} \,.
\end{equation}
If we can show that for some $\lambda_{\rm{min}} < 1$ we have $\textnormal{Tr} \, \sigma
C_{\lambda_{\rm{min}}} \geq 0$ for all $\sigma \in S$, $C_{\lambda_{\rm{min}}}$ is an entanglement
witness and therefore $\rho_{\rm{PPT}}$ and all states $\rho_\lambda$ with $\lambda_{\rm{min}} <
\lambda \leq 1$ are bound entangled (see Fig.~\ref{figgeneral}).

As an example we introduce the following family of three-parameter two-qutrit states (dimension $3
\times 3$ and $\mathbbm{1} := \mathbbm{1}_9$):
\begin{equation} \label{famstates}
    \rho_{\alpha,\beta,\gamma} \;:=\; \frac{1-\alpha -\beta -\gamma}{9}
    \mathbbm{1} + \alpha P_{00} + \frac{\beta}{2} \left( P_{10} +
    P_{20} \right) + \frac{\gamma}{3} \left( P_{01} +P_{11}+P_{21}
    \right)\,,
\end{equation}
where the parameters are constrained by the positivity requirement
$\rho_{\alpha,\beta,\gamma} \geq 0$. The states $P_{nm}$ are projectors onto maximally
entangled two-qutrit vector states \cite{narnhofer06} -- the \emph{Bell states}
\begin{equation} \label{bellstates}
    P_{nm} := (U_{nm} \otimes \mathbbm{1}) | \phi^+_3 \rangle \langle
    \phi^+_3 | (U_{nm}^\dag \otimes \mathbbm{1}) \,,
\end{equation}
where $| \phi^+_3 \rangle$ denotes the maximally entangled state $\left| \phi^+_d
\right\rangle = \frac{1}{\sqrt{d}} \,\sum_j \left| j \right\rangle \otimes \left| j
\right\rangle$ and $U_{nm}$ represent the \emph{Weyl operators} $U_{nm} = \sum_{k=0}^{d -
1} e^{\frac{2 \pi i}{d}\,kn} \,| k \rangle \langle (k+m) \,\textrm{mod}\,d|$ with $n,m =
0,1, \ldots ,d - 1 $ (here $d_1 = d_2 = d$).

The states \eqref{famstates} lie in the \emph{magic simplex} of two-qutrit states which
is the set of all two-qutrit states that can be written as a convex combination of the
projectors $P_{nm}$ \eqref{bellstates}. Viewing the indices $nm$ as points in a discrete
phase space the magic simplex reveals a high symmetry that makes its geometry much more
evident (see Refs.~\cite{baumgartner06,baumgartner07a, baumgartner07}).

For $\gamma = 0$ the geometry of the states \eqref{famstates} becomes rather simple. It is shown
in Ref.~\cite{baumgartner06} that all states of this two-parameter family are either NPT entangled
or separable, that is, all PPT states coincide with the separable states. Therefore we can easily
calculate the Hilbert-Schmidt measure with help of our entanglement witness \eqref{optoperator}.
\begin{figure}
  \includegraphics[width=0.5\textwidth]{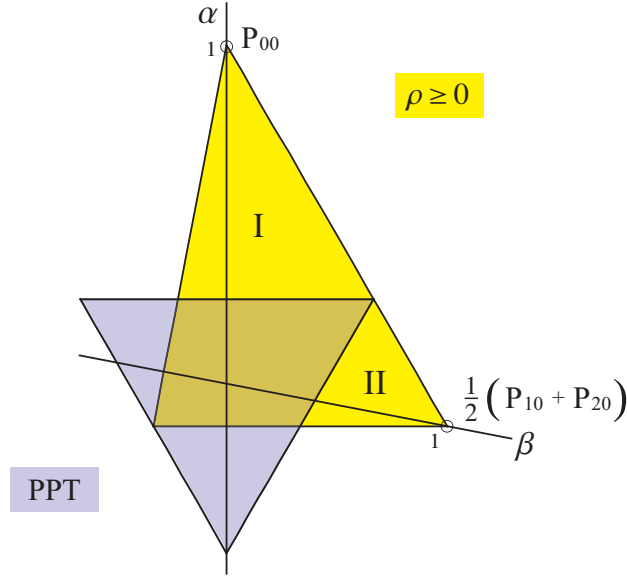}\\
  \caption{Illustration of states $\rho_{\alpha, \beta}$
  \eqref{famstates} ($\gamma = 0$) and their partial transposition. The regions I and II label
  the regions where the states are NPT entangled, they are PPT and separable in the overlap
  with the region of PPT points. The PPT points become semipositive under partial
  transposition.}\label{plotqutrit}
\end{figure}

In the two-dimensional picture (see Fig.~\ref{plotqutrit}) we can easily acquire the nearest
points on the border of the separable states -- here equal to the PPT states -- to the points in
the regions I and II that correspond to the entangled states: For region I the nearest point to
$(\alpha, \beta)$ is $(\frac{1}{4} + \frac{1}{8}\beta, \beta)$ which characterizes a state
$\tilde{\sigma}_\beta$ of the family \eqref{famstates}. For region II the nearest point is
$(\frac{1}{24} (-2+20\alpha+5\beta), \frac{1}{6} (2+4\alpha+\beta))$ which corresponds to a state
$\tilde{\sigma}_{\alpha, \beta}$. According to Ref.~\cite{bertlmann05} we can check if a proposed
separable state $\tilde{\sigma}$ is the nearest separable one to an entangled state $\rho$
(considering the geometry of \emph{all} states) by checking if the operator
\begin{equation} \label{checkoperator}
    C = \tilde{\sigma} - \rho - \langle \tilde{\sigma} , \tilde{\sigma}
    - \rho \rangle \mathbbm{1}_{\rm{D}}
\end{equation}
is an entanglement witness. To do so let us first state the following Lemma:
\begin{lem} \label{lemew}
    For any Hermitian operator $C$ of a bipartite Hilbert-Schmidt
    space of dimension $d \times d$ that is of the form
    \begin{equation} \label{lemqutritc}
        C \;=\; a \left( (d-1) \, \mathbbm{1}_{\rm{d^2}} \,+\, \sum_{n,m=0}^{d - 1} c_{nm}
            \,U_{nm} \otimes U_{-nm} \right), \quad a \in \mathbbm{R}^+, \
            c_{nm} \in \mathbbm{C}
    \end{equation}
    the expectation value for all separable states is positive,
    \begin{equation} \label{ewineqc}
        \langle \rho , C \rangle \,\geq\, 0 \quad \forall \, \rho \in S \,,
        \quad \mbox{\textrm{if}} \quad |c_{nm}| \,\leq\, 1 \quad \forall \, n,m \,.
    \end{equation}
\end{lem}
\noindent\emph{Proof.} Any bipartite separable state can be decomposed into Weyl
operators as
\begin{align} \label{sepweyl}
    \rho \;=\; & \sum_k p_k \ \frac{1}{d^2} \Big( \mathbbm{1_{\rm{d}}} \otimes
        \mathbbm{1_{\rm{d}}}
        \,+\, \sum_{n,m=0}^{d - 1} \sqrt{d-1}\, n_{nm}^k \,U_{nm} \otimes
        \mathbbm{1_{\rm{d}}}
        \,+\, \sum_{l,k=0}^{d - 1} \sqrt{d-1}\, m_{lk}^k \,\mathbbm{1_{\rm{d}}}
        \otimes U_{lk} \nonumber\\
    & + \sum_{n,m,l,k=0}^{d_1 - 1} (d-1)\, n_{nm}^k m_{lk}^k U_{nm} \otimes U_{lk} \Big) \,,
        \nonumber\\
    & n_{nm}^k, m_{lk}^k \in \mathbbm{C}\,, \;\quad \left| \vec{n}^k \right| \leq
        1\,,\quad \left| \vec{m}^k \right| \leq 1 \,, \quad \ p_k \geq 0, \; \sum_k p_k = 1\,,
\end{align}
where we define $\left| \vec{n}^k \right|^2 := \sum_{nm} n_{nm}^*n_{nm}$. We call a
decomposition \eqref{sepweyl} \emph{Bloch vector} form of the density matrix. Note that
for $d > 2$ not any vector with complex components represents a Bloch vector (a quantum state),
details can be found in Ref.~\cite{bertlmann07}.

Performing the trace we obtain (keeping notation $\rho^\dag$ formula \eqref{ew-sepstates}
becomes more evident)
\begin{equation}\label{ew-sepstates}
    \langle \rho , C \rangle \;=\; \textnormal{Tr} \rho^\dag \, C \;=\; \sum_k p_k
    \left( (d-1)a \left( 1 \,+\, \sum_{n,m} c_{nm} n_{nm}^{*k} m_{-nm}^{*k} \right) \right)\,,
\end{equation}
and using the restriction $|c_{nm}| \,\leq\, 1 \;\; \forall \, n,m $ we have
\begin{equation}
    \left| \sum_{n,m} c_{nm} n_{nm}^{*k} m_{-nm}^{*k} \right|
    \;\leq\; \sum_{n,m} |n_{nm}^k| |m_{-nm}^k| \;\leq\; 1 \,,
\end{equation}
and since the convex sum of positive terms stays positive we get $\langle \rho , C
\rangle \;\geq\; 0 \;\; \forall \rho \in S \,. \ \Box$\\

We set up the operators $C_I,C_{II}$ for the two regions,  respectively, according to
Eq.~\eqref{checkoperator} and write them in terms of Weyl operators normalized by $\|
\tilde{\sigma} - \rho \|$ for convenience,
\begin{equation}
    C_I= \frac{1}{6 \sqrt{2}} (2 \mathbbm{1}-U_1-U_2) \,, \quad
    C_{II}= \frac{1}{6 \sqrt{2}} (2 \mathbbm{1}+U_1-U_2) \,,
\end{equation}
where
\begin{align} \label{defu1u2}
    U_1 & := U_{01} \otimes U_{01} + U_{02} \otimes U_{02} + U_{11}
    \otimes U_{-11} + U_{12} \otimes U_{-12} + U_{21} \otimes U_{-21}
    + U_{22} \otimes U_{-22} \,, \nonumber\\
    U_2 & :=  U^I_2 + U^{II}_2  \qquad\mbox{with}\quad
    U^I_2 := U_{10} \otimes U_{-10} \,,\quad U^{II}_2 := U_{20} \otimes U_{-20} \,.
\end{align}
Both operators $C_I$ and $C_{II}$ satisfy Eq.~\eqref{ewineqc} of Lemma~\ref{lemew}. Due
to the construction of the operators, Eq.~\eqref{defentwitent} is satisfied for any
entangled states $\rho_{\alpha,\beta}$ in the regions I and II. Consequently $C_I$ and
$C_{II}$ are entanglement witnesses and therefore $\tilde{\sigma}_\beta$ and
$\tilde{\sigma}_{\alpha, \beta}$ are the nearest separable states $\sigma_{0; \, \beta}$
and $\sigma_{0; \, \alpha, \beta}$. The corresponding Hilbert-Schmidt measures of the
entangled two-parameter states $\rho_{\alpha,\beta}$ are
\begin{align}
    & D_I(\rho_{\alpha,\beta}^{\rm{ent}}) = \| \sigma_{0; \, \beta} -
    \rho_{\alpha,\beta}^{\rm{ent}} \| = \frac{2\sqrt{2}}{3}
    \left(\alpha - \frac{1}{4} - \frac{1}{8} \beta \right) \,. \\
    & D_{II}(\rho_{\alpha,\beta}^{\rm{ent}}) = \| \sigma_{0; \, \alpha, \beta} -
    \rho_{\alpha,\beta}^{\rm{ent}} \| = \frac{2\sqrt{2}}{6}
    \left(-\alpha - \frac{1}{2} + \frac{5}{4} \beta \right) \,.
\end{align}
Note that the measures can also be viewed as a maximal violation of the entanglement witness
inequality \eqref{defentwitsep}, as it is shown in detail in Refs.~\cite{bertlmann02,
bertlmann05}.

Another way to arrive at the nearest separable states for the two-parameter states is to
calculate the nearest PPT states with the method of Ref.~\cite{verstraete02} first and
then check if the gained states are separable. If we do so we obtain for the nearest PPT
states the states $\tilde{\sigma}_\beta$ and $\tilde{\sigma}_{\alpha, \beta}$ we have
found with our ``guess'' method, we know from Ref.~\cite{baumgartner06} that these states
are separable and therefore they have to be the nearest separable states.

Let us return to the family of three-parameter states $\rho_{\alpha, \beta, \gamma}$
\eqref{famstates}. For $\gamma \neq 0$ it is not trivial to find the nearest separable
states since the PPT states do not necessarily coincide with the separable states. But we
can use our geometric entanglement witness \eqref{cbe} to detect bound entanglement.

In Ref.~\cite{horodecki99c} the following one-parameter family of two-qutrit states was
introduced:
\begin{equation} \label{horstates}
    \rho_b \;=\; \frac{2}{7} \left| \phi_+^3 \right\rangle \left\langle
    \phi_+^3 \right| + \frac{b}{7} \, \sigma_+ + \frac{5 - b}{7} \,
    \sigma_- \,, \qquad 0 \leq b \leq 5 \,,
\end{equation}
where
\begin{align}
    \sigma_+ := \ & 1/3 \left( \left| 0 1 \right\rangle
        \left\langle 0 1 \right| + \left| 1 2 \right\rangle \left\langle
        1 2 \right| + \left| 2 0 \right\rangle \left\langle 2 0 \right|
        \right) \,, \nonumber\\
    \sigma_- := \ & 1/3 \left( \left| 1 0\right\rangle
        \left\langle 1 0 \right| + \left| 2 1 \right\rangle \left\langle
        2 1 \right| + \left| 0 2 \right\rangle \left\langle 0 2 \right|
    \right) \,.
\end{align}
Let us call this family of states \emph{Horodecki states}. Interestingly, the states
\eqref{horstates} are part of the three-parameter family \eqref{famstates}, namely
$\rho_b \equiv \rho_{\alpha, \beta, \gamma} \,, \;\;\mbox{with}\;\; \alpha = (6-b)/21 \,,
\;\beta = -2b/21 \,, \;\gamma = (5-2b)/7$ and thus lie in the magic simplex. Testing the
partial transposition we find that the Horodecki states \eqref{horstates} are NPT for $0
\leq b < 1$ and $4 < b \leq 5$ and PPT for $1 \leq b \leq 4$. In Ref.~\cite{horodecki99c}
it is shown that the states are separable for $2 \leq b \leq 3$ and bound entangled for
$3 < b \leq 4$.

We now want to pursue the following idea: Starting from a PPT and entangled -- bound
entangled -- Horodecki state $\rho_b^{\rm{BE}}$ we construct operators in the way of
Eq.~\eqref{cbe} and try to find bound entanglement on the line between $\rho_b^{\rm{BE}}$
and the maximally mixed state -- see Fig.~\ref{figgeneral}.

\begin{figure}
  \includegraphics[width=0.4\textwidth]{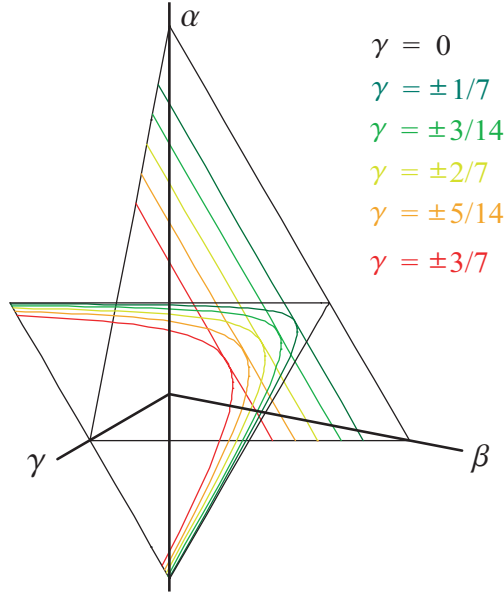}\\
  \caption{The states $\rho_{\alpha, \beta, \gamma}$ are depicted with slices of the
  three-dimensional parameter space for fixed values of $\gamma$. The curved borders of the PPT
  regions for $\gamma \neq 0$ are hyperbolae that result from intersecting a cone of PPT points
  with the planes of constant $\gamma$. The slices for positive and negative $\gamma$ overlap.
  All parameter axes are chosen non-orthogonal such that they become orthogonal to the boundary
  of the positivity region in order to reproduce the symmetry of the magic simplex.}
  \label{figallinone}
\end{figure}

For a geometric picture of the three-parameter family of states $\rho_{\alpha,\beta,\gamma}$ we
can fix values of $\gamma$ and draw two-dimensional slices (see Fig.~\ref{figallinone}).
Translated to values of $\gamma$ the states $\rho_b = \rho_{\alpha, \beta, \gamma}$ are
bound entangled for $-3/7 \leq \gamma < -1/7$ and in each slice of this region there is
exactly one point that corresponds to the bound entangled Horodecki state, namely $\left(
\alpha = (1+\gamma)/6 \,,\, \beta= (-5+7\gamma)/21 \right)$. The one-parameter line of
the states $\rho_b$ therefore cuts through the slices of fixed $\gamma$ and lies on the
boundary of the states \eqref{famstates}.

For a fixed $\gamma$ the operator $C_{\gamma, \lambda}$ \eqref{cbe} has the following
form:
\begin{align} \label{cbesimplex}
    & C_{\gamma, \lambda} \;=\; \rho_\lambda - \rho_b^{\rm{BE}} - \langle \rho_\lambda ,
    \rho_\lambda - \rho_b^{\rm{BE}} \rangle \mathbbm{1} \;=\; a (2 \,
    \mathbbm{1} + c_1 U_1 + c_2 U^I_2 + c^{\star}_2 U^{II}_2 ) \,, \nonumber\\
    & \mbox{with}\quad a = -\frac{1+3\gamma^2}{36}\lambda(\lambda-1), \;
    c_1 = - \frac{8}{7\lambda(1+ 3\gamma^2)}, \;
    c_2 = \frac{2(1-7 \sqrt{3} \gamma \, i)}{7 \lambda (1 + 3 \gamma^2)} \,,
\end{align}
and $\rho_\lambda$ represents the line of states
\begin{equation} \label{rholambda}
    \rho_\lambda \;=\; \lambda \rho_b^{\rm{BE}} + \frac{1-\lambda}{9} \mathbbm{1} \,.
\end{equation}
It is convenient to define functions $f_1(\gamma, \lambda) := |c_1|$ and $f_2(\gamma,
\lambda) := |c_2|\,$. We are interested to find the minimal $\lambda$ for a particular
$\gamma$ which is attained at $\max \{ f_1(\gamma, \lambda), f_2(\gamma, \lambda) \} =
1$, i.e. the minimal $\lambda$ such that $C_{\gamma,\lambda}$ \eqref{cbesimplex} is
necessarily an entanglement witness (see Lemma~\ref{lemew}). There exist values of
$\lambda$ with $\lambda < 1$ such that $\max \{ f_1(\gamma, \lambda), f_2(\gamma,
\lambda) \} < 1$ for $1/\sqrt{21} < |\gamma| \leq 3/7 $. That means we are able to detect
bound entangled states on lines \eqref{rholambda} between the bound entangled Horodecki
states and the maximally mixed state $\frac{1}{9}\mathbbm{1}$ until a value of $|\gamma|
= 1/\sqrt{21}$. These lines form a planar section of bound entangled states restricted by
$\lambda_{\rm{min}} < \lambda \leq 1$. As a ``side product'' we are able to detect bound
entanglement for the Horodecki states for $1 \leq b < \frac{1}{6}(15-\sqrt{21}) (\simeq
1.74)$ and $\frac{1}{6}(15+\sqrt{21})(\simeq 3.26) < b \leq 4$.

A detailed examination of the coefficient functions $f_1(\gamma, \lambda)$ and
$f_2(\gamma, \lambda)$ exhibits that for $1/\sqrt{21} < |\gamma| < \sqrt{5}/7$ we have
$f_1 > f_2$ and for $\sqrt{5}/7 < |\gamma| \leq 3/7$ we have $f_1 < f_2$ and equality
$f_1 = f_2$ at $\sqrt{5}/7$. The total minimum is $\lambda_{\rm{min}}^{\rm{tot}} = 7/8 =
0.875$, i.e. the minimal $\lambda$ that can be reached such that all states
$\rho_\lambda$ \eqref{rholambda} are detected to be bound entangled for
$\lambda_{\rm{min}}^{\rm{tot}} < \lambda \leq 1$. It is attained for $|\gamma| =
\sqrt{5}/7$ or $b=\frac{1}{2}(5-\sqrt{5}) \simeq 1.38$ and $b=\frac{1}{2}(5+\sqrt{5})
\simeq 3.62$.\\

Summarizing, the magic simplex contains three classes of states: NPT entangled states,
separable states but also PPT entangled or bound entangled states. We have found such
bound entangled states analytically in a quite large region of the magic simplex. Our method
works very well for the examples presented and is geometrically very intuitive. However,
we cannot detect in these examples the border between bound entangled and separable
states. The reason is that condition $|c_{nm}| \,\leq\, 1 \;\; \forall \, n,m $ in
Eq.~\eqref{ewineqc} is only sufficient for $C$ being an entanglement witness and not
necessary. There is generally still no operational method for detecting the nearest
separable states.

\bibliography{references}

\end{document}